\documentclass[11pt,twocolumn]{article}

\usepackage{graphicx}
\usepackage{dcolumn}
\usepackage{authblk}
\usepackage{amsmath}
\usepackage{amssymb}
\usepackage[a4paper, total={7.2in, 10in}]{geometry}

\newcommand{\dgr}{^{\circ}}

\begin{document}

\title{Surface Bound States in the Continuum in Dyakonov Structures}
\date{}
\author[1]{Samyobrata Mukherjee}
\author[1, 2, *]{David Artigas}%
\author[1,2]{Lluís Torner}
\affil[1]{ICFO-Institut de Ciències Fotòniques, The Barcelona Institute of Science and Technology, 08860 Castelldefels (Barcelona), Spain}
\affil[2]{Department of Signal Theory and Communications, Universitat Politècnica de Catalunya, 08034 Barcelona, Spain}
\affil[*]{david.artigas@icfo.eu}

\twocolumn[
  \begin{@twocolumnfalse}
    \maketitle
    \begin{abstract}
    Surface bound states in the continuum (SBICs) have been found to occur in diverse settings, but so far always at the interface of non-homogeneous media, such as discrete lattices or periodic systems. Here we show that they can also exist at the interface of homogeneous media, resulting in unique SBICs. Specifically, we found that, contrary to general belief, leaky Dyakonov states exist at the interface between materials that exhibit opposite signs of anisotropy. In addition, properly breaking the anisotropy-symmetry leads to the formation of both guided states and also SBICs embedded within the continuum.  A direct implication of our finding is the possibility to create SBICs and Dyakonov states in a whole new class of materials and metamaterials.
    \end{abstract}
    \vspace{0.5cm}
  \end{@twocolumnfalse}
  ]

\section{Introduction}
Bound states in the continuum (BICs) are radiationless modes that are embedded in the part of the spectrum corresponding to radiating modes. First discovered in the context of quantum mechanics \cite{Neuman1929, Stillinger1975}, BICs are now understood as a general wave phenomenon \cite{Hsu2016, Joseph2021}. The existence of photonic BICs was first theoretically predicted in parallel arrays of dielectric gratings and cylinders \cite{Marinica2008} and in photonic crystals of dielectric rods with defects \cite{Bulgakov2008}. This was followed by landmark experimental demonstrations of photonic BICs in an array of waveguides with defects \cite{Plotnik2011} and in a photonic crystal slab \cite{Hsu2013}. This led to an explosion of interest in the study of photonic BICs in various geometries such as layered nanospheres \cite{Monticone2014}, anisotropic waveguides \cite{Gomis-Bresco2017}, 1D photonic crystals with anisotropic defect layers \cite{Timofeev2018, Pankin2020}, epsilon-near-zero metamaterials \cite{Minkov2018}, plasmonic systems \cite{Azzam2018, Sun2021}, all-dielectric and plasmonic metasurfaces \cite{Fan2019, Liang2020}, $PT$-symmetric systems \cite{Longhi2014, Kartashov2018, Song2020}, tailored photonic crystal environments \cite{Cerjan2019, Cerjan2021} and other periodic systems \cite{Bulgakov2014b, Bulgakov2015, Bulgakov2017, Bulgakov2017a}. Several potential applications of photonic BICs have also been suggested \cite{Kodigala2017, Romano2018, Carletti2018, Hayran2021}. The photonic BICs in all these geometries involve localising light in a waveguide \cite{Bulgakov2008, Gomis-Bresco2017} or a resonator \cite{KoshelevReview} by cancelling the coupling of the light to an available radiation channel by exploiting mechanisms such as symmetry protection or destructive interference. Another class of photonic BICs, known as surface BICs (SBICs), exist where the light is localised at an interface between two unlike materials. SBICs have been reported at the interfaces of finite 1D arrays of waveguides \cite{Molina2012, Weimann2013, Corrielli2013, Gallo2014} and photonic crystals \cite{Hsu2013a, Hu2017, Chai2020, Tasolamprou2020}. However, all SBICs reported so far involve discrete lattices or photonic crystals, whose characteristic geometric dimensions restrict the existence of SBICs to specific wavelengths.

In this paper we report the existence of SBICs at the interface between two semi-infinite, fully homogeneous media. Specifically, we address the interface between two anisotropic uniaxial materials with opposite signs of birefringence. Interfaces involving birefringent materials are known to support hybrid, full-vector Dyakonov surface waves (DSWs) when certain conditions are met \cite{Dyakonov1988, Averkiev1990, Takayama2008, Osamu2009}. To date, all lossless DSWs are known to occur in structures made of positive birefringent materials, while only generalized states arising in non-Hermitian or truncated structures have been found to exist elsewhere  \cite{Mackay2019a,Repan2020,Chermoshentsev2021}. However, here we discovered that material interfaces between media with opposite sign of anisotropy and, importantly, where the anisotropy-symmetry is broken \cite{Averkiev1990,Mukherjee2018}, may also support lossless DSWs. In addition, in such structures radiation channels that couple the states otherwise fully localized at the interface with the continuum arise, thus creating leaky surface states. Nevertheless, we also discovered that, under suitable conditions, the radiation channel of such leaky surface states can be suppressed completely, transforming them into  anisotropy-induced BICs \cite{Gomis-Bresco2017}, which, therefore, are lossless SBICs. To the best of our knowledge, this is the first known example of surface bound states in the continuum  supported by the interface between two homogeneous media.

\section{Theoretical Formulation}

\begin{figure}[t!]
    \centering
    \includegraphics[width=0.9\linewidth]{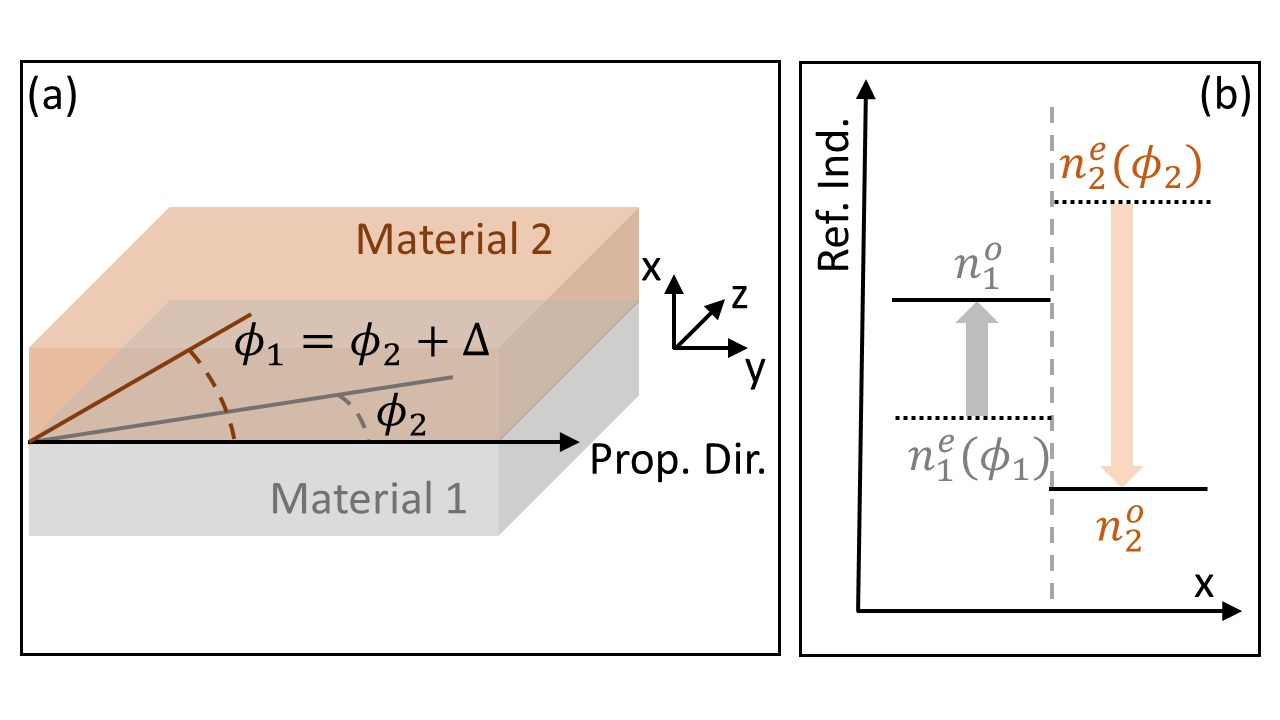}
    \caption{(a) Geometry comprising two semi-infinite uniaxial materials with their interface at $x=0$. A schematic of the optic axes layout is shown. (b) Schematic of the refractive indices of the system. The dashed gray line shows the interface. The solid black lines show the ordinary indices which are constant while the dotted black lines indicate the extraordinary indices which vary with optic axis orientation in the range shown by the colored arrows.}
    \label{fig:1}
\end{figure}
We address the planar interface located at $x=0$ where the negative and positive $x$-half spaces are occupied by lossless, uniaxial, dielectric materials with negative (material 1) and positive (material 2) birefringence, respectively.  We study bound solutions of the full Maxwell equations for this structure in different configurations. Light propagation is along the $y$ direction, so $k_y$ is the propagation constant and $k_0$ the free space wavenumber. The optic axes of materials 1 and 2 lie in the interface plane and make angles $\phi_1$ and $\phi_2$ with the direction of light propagation. The layout is shown in Fig.~\ref{fig:1}(a). The structure is then anisotropy-symmetric when $\phi_1=\phi_2$ while introducing an offset $\Delta=\phi_1-\phi_2$ results in azimuthal, or weakly, anisotropy-symmetry breaking \cite{Mukherjee2018}. The permittivity tensor in material $i$ when $\phi_i=0$ is $\hat{\epsilon} = diag(\epsilon^o_i,\epsilon^e_i,\epsilon^o_i)$, where $\epsilon^o_i$ and $\epsilon^e_i$ are its principal values. The ordinary refractive index is constant and is given by $n^o_i=\sqrt{\epsilon^o_i}$, whereas the extraordinary refractive index varies as a function of the optic axis orientation $n^e_i(\phi)= \sqrt{{\epsilon^e_i}}{(\sin^2\phi_i + \frac{\epsilon^e_i}{\epsilon^o_i}\cos^2\phi_i)^{-1/2}}$. For a given value of the mode effective index $N=k_y/k_0$, four basis waves propagating in a uniaxial material (two ordinary and two extraordinary) can be obtained from an eigenvalue equation \cite{Hodgkinson_Book, Mukherjee2018}.  One of each kind propagates in the $+x$ direction while the other propagates in the $-x$ direction. The transverse component of the ordinary, $\kappa^o=k_x^o/k_0$, and extraordinary, $\kappa^e=k_x^e/k_0$ normalized wavevectors, are obtained as the eigenvalues 
\begin{equation}
\begin{split}
    \kappa^o & = \pm\sqrt{\epsilon^o-N^2}, \\
    \kappa^e(\phi) & = \pm\sqrt{\epsilon^e - N^2\left( \sin^2\phi + \frac{\epsilon^e}{\epsilon^o}\cos^2\phi \right)},
\end{split}
\label{eigenval}
\end{equation}
with the corresponding eigenvectors
\begin{equation}
    \vec{F}^o=\begin{bmatrix}
    \kappa^o\sin{\phi} \\ \epsilon^o\sin{\phi} \\ -\kappa^o\cos{\phi} \\ (\kappa^o)^2\cos{\phi}
    \end{bmatrix},\ \ \ \vec{F}_e=\begin{bmatrix} 
    (\kappa^o)^2\cos{\phi} \\ \epsilon^o\kappa^e\cos{\phi} \\ \epsilon^o\sin{\phi} \\ -\epsilon^o\kappa^e\sin{\phi}
    \end{bmatrix},
\end{equation}
where the four rows of $\vec{F}_o$ and $\vec{F}_e$ correspond to the tangential electric and magnetic field components $E_y, z_oH_z, E_z$ and $z_0H_y$, where $z_0$ is the vacuum impedance. One ordinary and one extraordinary out of the four waves, have to be selected in each media. For a standard DSW, the effective index is real with  $N>n^o,n^e(\phi)$, and the basis waves decaying exponentially perpendicular to the interface are selected. For the leaky DSW, the effective index is complex $N$, with $\Re(N)$ lesser than one of the refractive indices of the system. The basis wave corresponding to that index is defined as the radiation channel via which the mode couples to the continuum where $\Im(N)$ approximates the wave radiation loss. Due to flux considerations, the radiation channel basis wave must grow exponentially away from the interface \cite{Hu2009}. In this letter, leaky DSWs feature $\Re(N)<n^o_1$, thus the ordinary wave in the negative uniaxial media is the radiation channel.

Boundary conditions for the tangential field components at the interface between material $1$ and $2$ yield the homogeneous set of linear equations
\begin{equation}
    a_1^{o}\vec{F}_1^{o} + a_1^{e}\vec{F}_1^{e} = a_2^{o}\vec{F}_2^{o} + a_2^{e}\vec{F}_2^{e},
    \label{eq:3}
\end{equation}
where $a_{i}^{o/e}$ is the amplitude of the corresponding basis wave. Writing eq. (\ref{eq:3}) in a matrix form, $\hat{R}\vec{a}=0$, with $\hat{R}= \begin{bmatrix} \vec{F}_1^{o} & \vec{F}_1^{e} & -\vec{F}_2^{o} & - \vec{F}_2^{e} \end{bmatrix}$ and  $\vec{a}=\begin{bmatrix} a^o_1 & a^e_1 & a^o_2 & a^e_2\end{bmatrix}^T$, being a $4\times4$ matrix and a $4\times1$ column vector, respectively, the requirement for non-trivial solution, $|\hat{R}|=0$, yields the dispersion equation
\begin{equation}
\begin{split}
    2 \epsilon_{2}^{o} \epsilon_{1}^{o} \kappa^{o}_{1} \kappa^{o}_{1} \left(\kappa^{e}_{1} - \kappa^{o}_{2}\right) \left(\kappa^{e}_{1} - \kappa^{o}_{1}\right) \sin{\phi_{2} } \sin{\phi_{1} } \cos{\phi_{2} } \cos{\phi_{1}} \\
    + \kappa^{o}_{2} \kappa^{o}_{1} \left( \kappa^{o}_{1}- \kappa^{o}_{2}\right) \left( \epsilon_{1}^{o} \kappa^{e}_{1} \left(\kappa^{o}_{2}\right)^{2}- \epsilon_{2}^{o} \kappa^{e}_{2} \left(\kappa^{o}_{1}\right)^{2}\right) \cos^{2}{\phi_{2} } \cos^{2}{\phi_{1}}\\
    + \epsilon_{2}^{o} \kappa^{o}_{1} \left(\kappa^{e}_{2} - \kappa^{o}_{1}\right) \left(\epsilon_{2}^{o} \left(\kappa^{o}_{1}\right)^{2} - \epsilon_{1}^{o} \kappa^{e}_{1} \kappa^{o}_{2}\right) \sin^{2}{\phi_{2} } \cos^{2}{\phi_{1} } \\
    + \epsilon_{1}^{o} \kappa^{o}_{2} \left(\kappa^{e}_{1} - \kappa^{o}_{2}\right) \left( \epsilon_{1}^{o} \left(\kappa^{o}_{2}\right)^{2}- \epsilon_{2}^{o} \kappa^{e}_{2} \kappa^{o}_{1}\right) \sin^{2}{\phi_{1} } \cos^{2}{\phi_{2} } \\
    + \epsilon_{2}^{o} \epsilon_{1}^{o} \left(\kappa^{e}_{2} - \kappa^{e}_{1}\right) \left(\epsilon_{2}^{o} \kappa^{o}_{1} - \epsilon_{1}^{o} \kappa^{o}_{2}\right) \sin^{2}{\phi_{2} } \sin^{2}{\phi_{1}}=0.
    \end{split}
    \label{eq:4}
\end{equation}
The transcendental Eq. (\ref{eq:4}) is solved numerically to obtain the effective index $N$ for the range of values $\phi_1$ and $\phi_2$ over which the solution exists, resulting in the narrow range of propagation directions typical in DSWs. Purely real solutions of eq. (\ref{eq:4}) correspond to standard guided DSWs whereas complex solutions of eq. (\ref{eq:4}) correspond to leaky DSWs. SBICs are embedded on the leaky DSWs and exist when $\Im (N)=0$. A necessary but insufficient condition for the existence of surface waves in this geometry is an overlap between the extraordinary indices of the two materials, i.e., in terms of the permittivities: 
\begin{equation}
    [\epsilon_1^o>\epsilon_2^o]\ \land \ [\epsilon_1^e<\epsilon_2^e].
\end{equation}

\section{Results and Discussion}
Without loss of generality, we consider a structure with a negative uniaxial material with $n^{o}_1=\sqrt{\epsilon^o_1}=1.80$ and $n_1^e(\phi_1=90\dgr)=\sqrt{\epsilon^e_1}=1.40$, and a positive uniaxial material with  $n^{o}_2=\sqrt{\epsilon^o_2}=1.25$ and $n_2^e(\phi_2=90\dgr)=\sqrt{\epsilon^e_2}=2$.  A schematic of the refractive indices is shown in Fig. \ref{fig:1}(b).

\subsection{Leaky DSWs}
\begin{figure}[t!]
    \centering
    \includegraphics[width=0.9\linewidth]{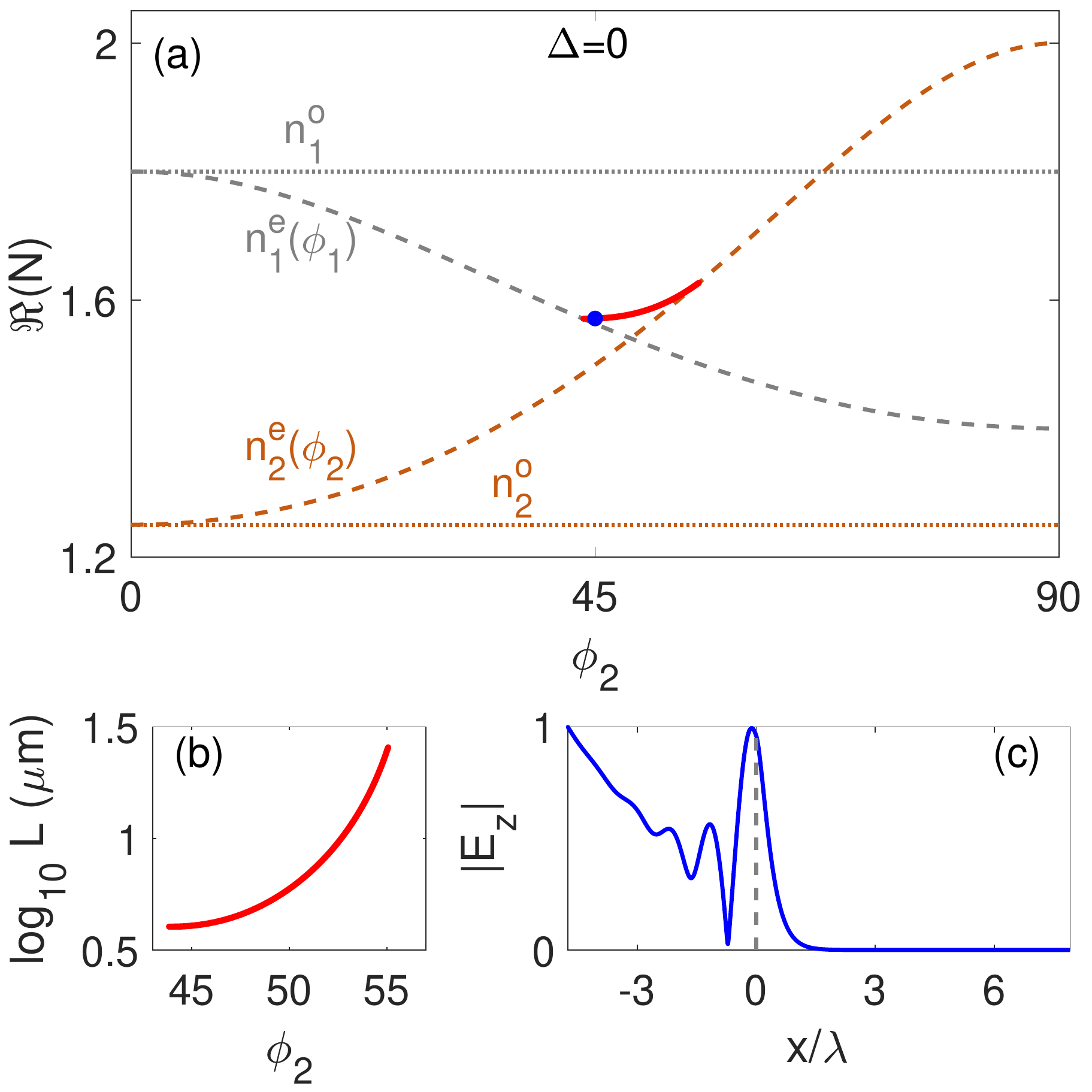}
    \caption{ (a) Mode propagation constant $\Re(N)$ of DSW (red) as a function of $\phi_2$ when $\Delta=0\dgr$ and only leaky DSWs exist. The ordinary (dotted lines) and extraordinary (dashed lines) refractive indices of material 1 and 2 are plotted in gray and brown, respectively. (b) $1/e$ propagation length $L$ (log scale) for the leaky DSW with $\lambda=0.632\ \mu m$. (c) Transverse profile of the field $|E_z|$ at $\phi_2=45\dgr$ (blue dot in (a)). The dashed gray line at $x/\lambda=0$ indicates the interface.}
    \label{fig:2_leaky}
\end{figure}

Figure \ref{fig:2_leaky}(a) shows the propagation constant of the DSW when the optic axes in the two materials are aligned ($\Delta=0\dgr$),  corresponding to a structure that maintains anisotropy-symmetry. A surface wave exists in the vicinity of the point where $n_1^e(\phi_1)= n_2^e(\phi_2)$ and cuts off at the points where $\Re(N)=n_1^e(\phi_1)$ and $\Re(N)=n_2^e(\phi_2)$. $\Re(N)$ is greater than $n_1^e(\phi_1)$, $n_2^e(\phi_2)$ and $n_2^o$ and the corresponding three basis waves decay exponentially away from the interface. However, $\Re(N)<n_1^o$, and therefore, the ordinary basis wave in material 1 acts as the radiation channel via which the leaky mode couples to the continuum. Therefore, with $\Delta=0\dgr$, and for small values of azimuthal anisotropy-symmetry breaking, the surface wave is a leaky DSW for all values of $\phi_2$ where the solution exists. However, there are no SBICs embedded on the leaky DSW in this configuration.

Figure \ref{fig:2_leaky}(b) shows the logarithm of $L$, defined as the $1/e$ propagation length for the leaky DSW for wavelength $\lambda=0.632\ \mu m$. While $L$ is larger at larger  values of $\phi_2$, its value for the leaky DSW is generally small ($\sim 10\ \mu m$). Figure \ref{fig:2_leaky}(c) shows the field amplitude $E_z$ along the transverse axis $x$ for the leaky DSW at $\phi_2=45\dgr$. Though there is radiation away from the DSW, the field is localized at the interface. Since $N$ is complex for leaky DSWs, the $x$-component of the radiation channel's normalized wavevector, $\kappa^o_1$, is also complex, with $\Im(\kappa^o_1)$ and $\Re(\kappa^o_1)$ resulting, respectively, in the exponential growth and the oscillations in the transverse field profile leaking from the DSW, as seen in Fig. \ref{fig:2_leaky}(c).

\subsection{Standard guided DSWs}

\begin{figure}[t!]
    \centering
    \includegraphics[width=0.9\linewidth]{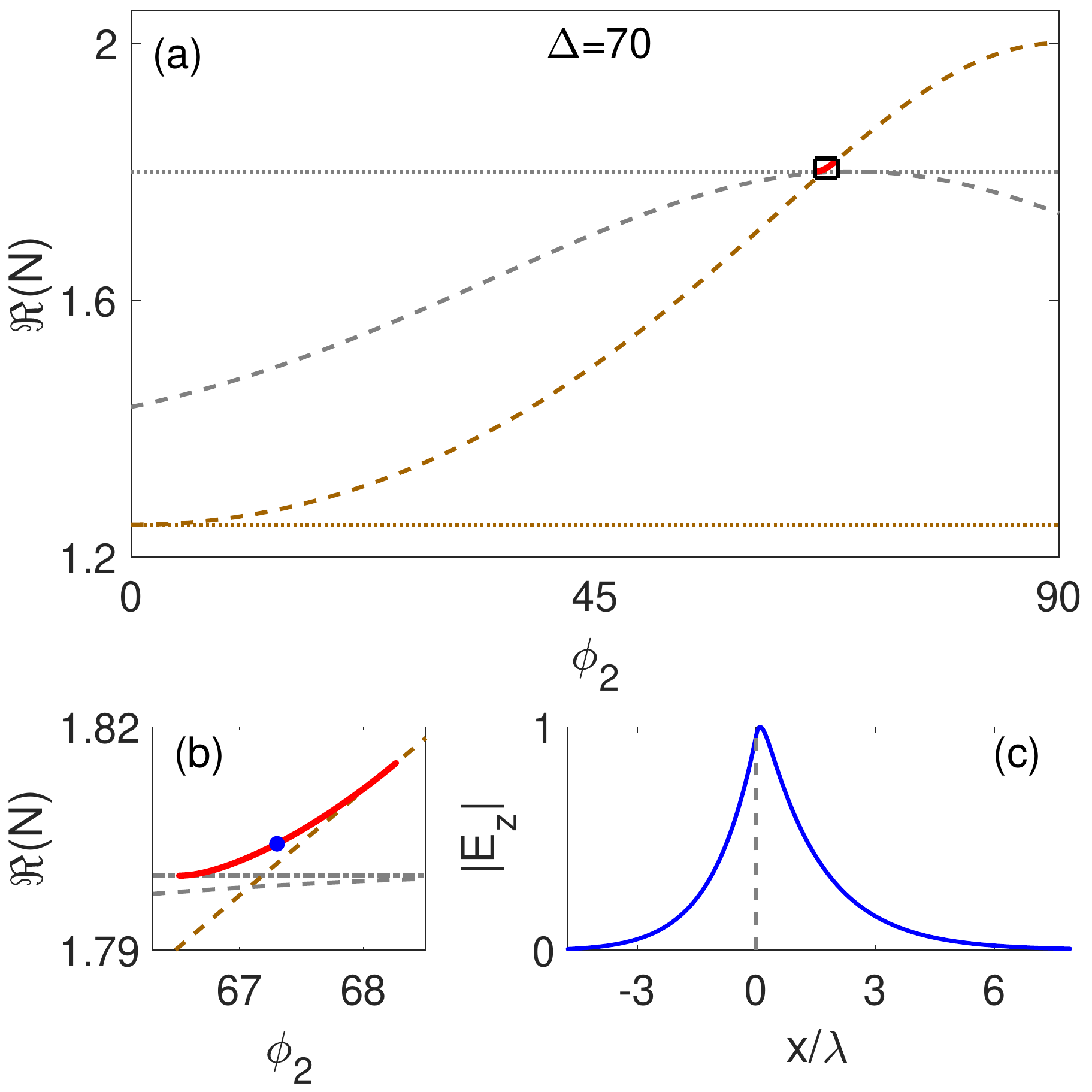}
    \caption{(a) Same as Fig. \ref{fig:2_leaky}(a) but when $\Delta=-70\dgr$ and only guided DSWs exist. (b) Magnification of the area in the black box in (a). (c) Transverse profile of the field $|E_z|$ at $\phi_2=67.3\dgr$ (blue dot in (b)).}
    \label{fig:3_guided}
\end{figure}
Figure \ref{fig:3_guided}(a) shows the surface wave supported by a structure for an amount of azimuthal anisotropy-asymmetry given by $\Delta=-70\dgr$. The surface wave solution in this structure cuts off at $N=n_1^o$ on one side and at $N=n_2^e(\phi_2)$ on the other side. Throughout all the range of existence, $N$ is purely real and is greater than all four refractive indices of the structure. Therefore, this structure supports standard, guided DSWs that are made up of four basis waves which decay exponentially away from the interface, as Fig.~\ref{fig:3_guided}(c) shows at $\phi_2=67.3\dgr$. As is typical in DSWs, when compared with propagation directions near the DSW cutoff $N=n_1^o$, the field of the DSW remains primarily in material 1, while traversing away from this point towards the cutoff $N=n_2^e(\phi_2)$ results in the majority of the field moving to material 2 \cite{Takayama2008}. The structure exhibits the standard challenge regarding excitation of DSWs, i.e., a narrow range of allowed propagation angles $\phi_2$ (red line in Fig.~\ref{fig:3_guided}(a)). However, this range is much broader when leaky DSWs are considered (red line in Fig.~\ref{fig:2_leaky}(a)).

There have been a few reports of the existence of surface waves at interfaces involving negative uniaxial materials, however these proposals contain added constraints such as dissipative media \cite{Mackay2019a}, birefringent metals \cite{Repan2020}, or a finite interface \cite{Chermoshentsev2021}. This is the first report of the existence of a guided DSW at an infinite planar interface between a positive uniaxial material and negative uniaxial material.

\subsection{SBIC embedded on leaky branches of DSWs}
\begin{figure}[t!]
    \centering
    \includegraphics[width=0.9\linewidth]{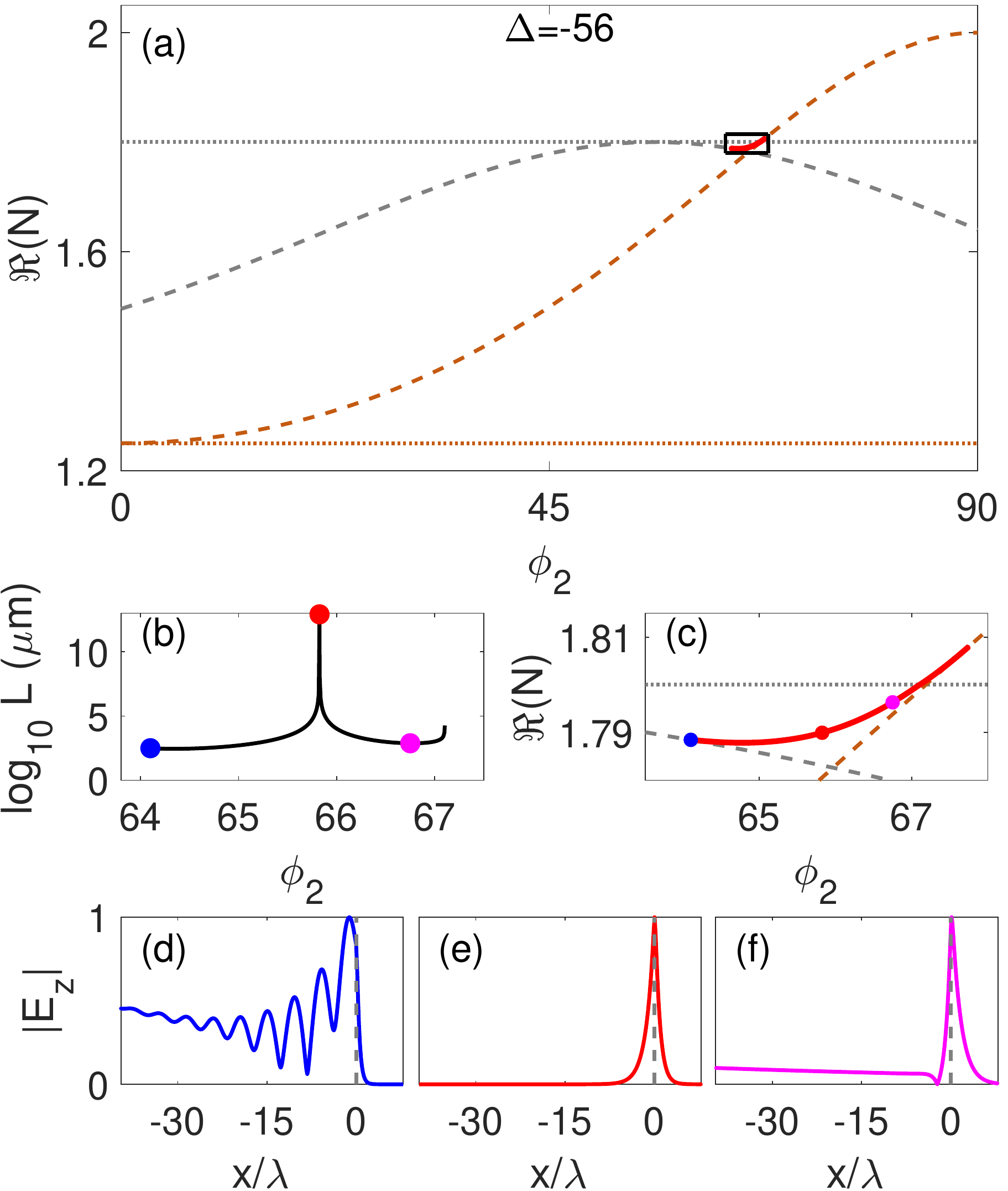}
    \caption{(a) Same as Fig. \ref{fig:2_leaky}(a) but for $\Delta=-56\dgr$. (b) $1/e$ propagation length $L$ (log scale), of the leaky DSW in (a) for $\lambda=0.632 \ \mu m$. $L$ diverges at the DSW BIC (red dot). (c) Magnification of the area in the black box in (a). The coloured markers in (b) and (c) show the values of $\phi_2$ where we plot the normalized transverse profile $|E_z|$ for a (d) leaky DSW at $\phi_2=64.1\dgr$, (e) the BIC at $\phi_2=65.82\dgr$, and (f) a leaky DSW at $\phi_2=66.75\dgr$.}
    \label{fig:4_mixed}
\end{figure}

For a range of values of $\Delta$ between the two previous cases  with varying amounts of azimuthal anisotropy-asymmetry, the structure supports both purely guided and leaky DSWs, as shown in Fig. \ref{fig:4_mixed}(a) for $\Delta=-56\dgr$. The leaky DSW exists at values of $\phi_2$ where $\Re(N) \leq n_1^o$, with the ordinary wave in material 1 serving as the radiation channel. The transition to purely guided DSWs occurs for higher values of $\phi_2$, where $N$ is greater than all the refractive indices in the system and therefore it is purely real.

We study the conditions for cancellation of the radiation channel for the leaky DSW using an auxiliary condition by setting the radiation channel amplitude in eq. (\ref{eq:3}) to zero, $a^o_1=0$, \cite{Gomis-Bresco2017, Mukherjee2021}. The auxiliary condition yields four auxiliary equations, one of which is
\begin{equation}
    \begin{split}
        \epsilon^o_2 \epsilon^o_1 \kappa^o_2 \left(- \kappa^e_2 + \kappa^e_1\right) \sin^{2}{\left(\phi_{c} \right)} \sin{\left(\phi_{s} \right)} +\\ \epsilon^o_2 \kappa^o_2 \left(\kappa^o_1\right)^{2} \left(- \kappa^e_2 + \kappa^o_2\right) \sin{\left(\phi_{c} \right)} \cos{\left(\phi_{c} \right)} \cos{\left(\phi_{s} \right)}\\ + \epsilon^o_1 \left(\kappa^o_2\right)^{3} \left(\kappa^e_1 - \kappa^o_2\right) \sin{\left(\phi_{s} \right)} \cos^{2}{\left(\phi_{c} \right)}=0.
    \end{split}
    \label{auxeq}
\end{equation}
The SBIC exists only when the solutions of the dispersion equation and all four auxiliary equations coincide, resulting in $\phi_2^{BIC}=65.82\dgr$.  We calculate $L$, the $1/e$ propagation length from $\Im(N)$ over the range of the leaky DSW (Fig. \ref{fig:4_mixed}(b)), showing that $L$ diverges at the angle found before, $\phi_2^{BIC}$, resulting in a SBIC where the leakage is canceled. Fig. \ref{fig:4_mixed}(e) shows the field profile $|E_z|$ for the SBIC, showing that despite being embedded in the leaky part of the DSW with $\Re(N)<n_1^o$, the field decays exponentially away from the interface. This contrasts with the field profile for two leaky DSWs on either side of the SBIC, shown in Fig. \ref{fig:4_mixed}(d) and (f), at $\phi_2=64.1\dgr$ and $\phi_2=66.75\dgr$, respectively. Both figures show the characteristic radiation into the substrate via the radiation channel. 

\subsection{Cut-offs in $\Delta$}
The amount of azimuthal anisotropy-asymmetry given by $\Delta$ has a dramatic impact on the properties of the DSW and the existence of SBICs. We explore this effect in Fig. \ref{fig:5}(a), where the solid black lines show the upper and lower bounds of $\Re(N)$ as a function of $\Delta$.

\subsubsection{Cutoffs for standard guided DSW}
\begin{figure*}[t!]
    \centering
    \includegraphics[width=0.8\linewidth]{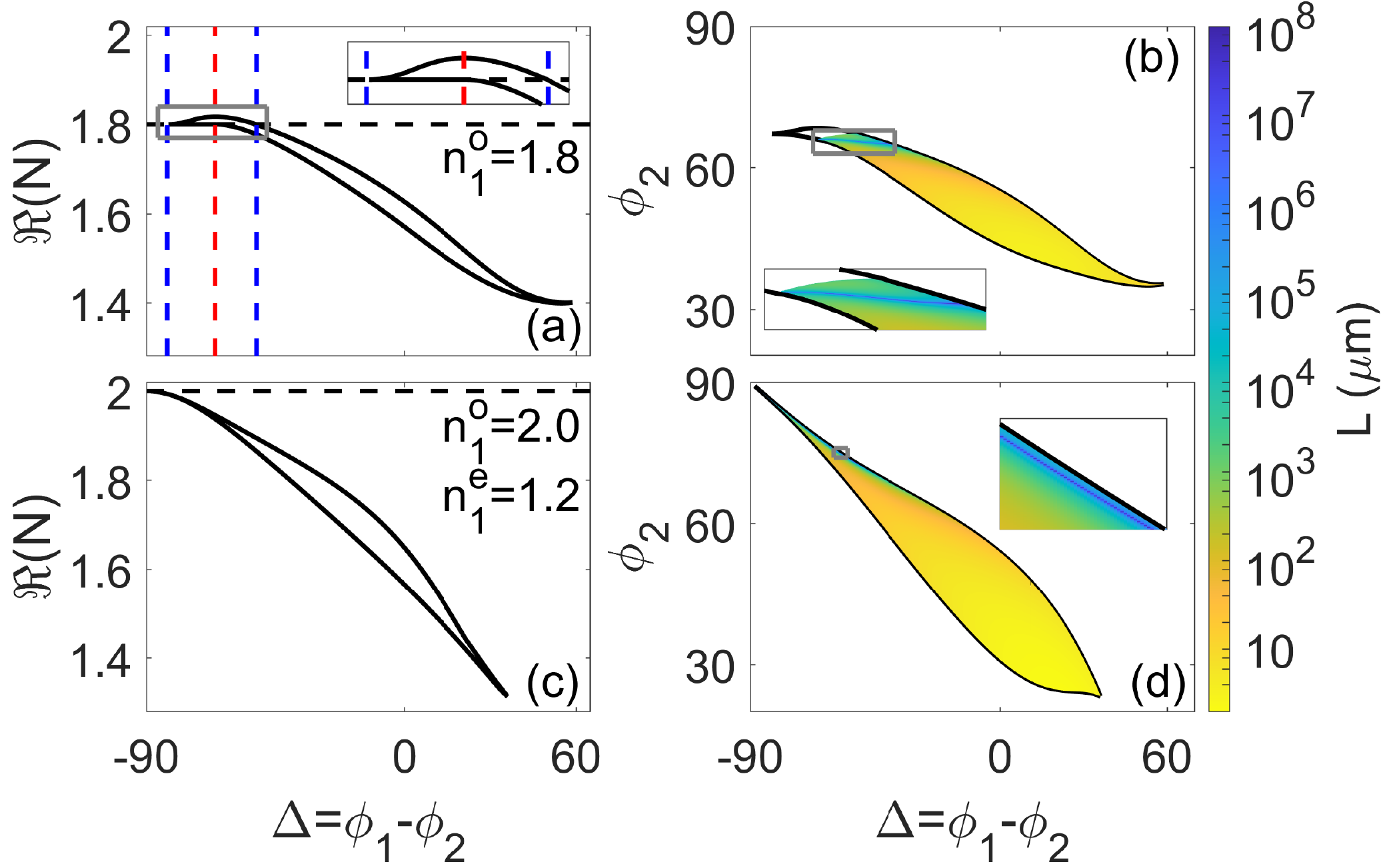}
    \caption{(a) Range of $\Re(N)$ as a function of $\Delta$ for the structure studied in Figs. \ref{fig:2_leaky}, \ref{fig:3_guided}, and \ref{fig:4_mixed}. The dashed black line shows the value of $n^o_1=1.8$ below which only leaky DSWs can exist. The dashed blue lines show the point where guided DSWs start ($\Delta_{cm}$) and then cease ($\Delta_{cM}$) to exist. The dashed red line shows the value of $\Delta_{cL}$ where the leaky DSW first appears. (b) Range of $\phi_2$ as a function of $\Delta$. The color indicates the $1/e$ propagation length, $L$ for leaky DSWs for $\lambda=0.632\ \mu m$. The insets shows the line of BICs. (c-d) Same as (a-b) but for $n^o_1=2, n^e_1=1.2$.}
    \label{fig:5}
\end{figure*}

We see that the guided DSW solution only exists for a finite range of values of $\Delta$, those where $\Re(N)>n_1^o=1.8$ (between the two blue dashed lines). As we show in Figs. \ref{fig:3_guided}(a) and \ref{fig:4_mixed}(a), the guided DSW cuts off at the points where $N=n_1^o$ ( $\implies\kappa^o_1=0$) and $N=n_2^e(\phi_2)$ ($\implies\kappa^e_2=0$). Making use of $n_1^o=n_2^e(\phi_2)$, we obtain the cutoff angle $\phi_{2c}$ as 
\begin{equation}
     \sin{\phi_{2c}}=\pm\sqrt{\frac{\epsilon_2^e (\epsilon_2^o - \epsilon_1^o)} {\epsilon_1^o (\epsilon_2^o - \epsilon_2^e)}},
    \label{eq:5}
\end{equation}
and then using $\kappa^o_1=0$ and $\kappa^e_2=0$ in the dispersion equation (eq. (\ref{eq:4})), together with equation (\ref{eq:5}), we find the cutoff angle
\begin{equation}
    \sin{\phi_{1c}}=\pm\frac{(\epsilon_1^o - \epsilon_2^e)}{\epsilon_1^o}
    \sqrt{\frac{(\epsilon_2^o - \epsilon_1^o)} {(\epsilon_1^e - \epsilon_1^o)}}.
    \label{eq:6}
\end{equation}
Thus, the cutoff in terms of azimuthal anisotropy-asymmetry can be calculated as $\Delta_{c}=\phi_{1c}-\phi_{2c}$. Since $N$ is a function of $\phi_2$, the sign selected in eq. (\ref{eq:6}), determines whether the leading or trailing edge of $N$ passes through the point $n_1^o=n_2^e(\phi_2)$. The positive (negative) sign gives the value of $\Delta_{cm}$ ($\Delta_{cM}$) where pure guided DSWs start (cease) to occur (dashed blue lines in Fig. \ref{fig:5}(a)). 

\subsubsection{Cutoff for leaky DSW}
It is useful to determine  another cutoff in terms azimuthal anisotropy-asymmetry, $\Delta_{cL}$, where leaky DSWs start existing (dashed red line in Fig. \ref{fig:5}(a)) since SBICs are embedded in the leaky DSW. Leaky DSWs occur when $\Re(N)< n_1^o$. They start to exist at $N=n_1^o=n_1^e(\phi_1)$ which implies $\kappa^o_1=\kappa^e_1=0$. This condition can only be fulfilled when $\phi_1=0\dgr\implies\phi_2= -\Delta$. Applying these conditions to eq. (\ref{eq:4}), we find that $\Delta_{cL}$ obeys the equation  
\begin{equation}
    \tan^2{\Delta_{cL}} +\frac{(\epsilon_2^o - \epsilon_1^o)^{3/2}}{\epsilon_2^o \sqrt{\epsilon_2^e - \epsilon_1^o\left( \sin^2{\Delta_{cL}} + \frac{\epsilon_2^e}{\epsilon_2^o}\cos^2{\Delta_{cL}} \right)}}=0,
    \label{eq:7}
\end{equation}
which can be readily solved numerically.

Equations (\ref{eq:5}-\ref{eq:7}) determine the existence conditions for DSWs in terms of the azimuthal anisotropy-asymmetry $\Delta$. Guided DSWs only exist at negative values of $\Delta$, with $\Delta>\Delta_{cm}$. As $\Delta$ increases, leaky DSWs start to appear at $\Delta>\Delta_{cL}$, co-existing with guided DSWs, and when $\Delta>\Delta_{cM}$, only leaky DSWs exist. In all cases DSWs propagate in a range of values of $\phi_2^{DSW}$, as shown in Fig. \ref{fig:5}(b), where the color shows the $1/e$ propagation distance $L$. The leaky DSW becomes more radiative (low values of $L$) as $\Delta$ is increased. Note that, while $\phi_2^{DSW}$ is narrow for guided DSWs, reaching the larger value $\phi_2^{DSW} \sim 4 \dgr$ for $\Delta=\Delta_{cL}$,  it increases for leaky DSWs to $\phi_2^{DSW} \sim 22 \dgr$ near $\Delta=0$. Note that while, by and large, the usually narrow range of allowed propagation angles for standard DSWs to exist is an outstanding challenge for their experimental generation \cite{Osamu2009}, the larger range of propagation directions of leaky DSWs facilitates their excitation.

\subsection{Variation of the constitutive parameters}

Equation \ref{eq:7} provides the starting point beyond which SBICs, embedded in leaky DSWs, can exist in terms of the azimuthal anisotropy-asymmetry, $\Delta$. Indeed, the structure supports a line of SBICs for $\Delta>\Delta_{cL}$, shown by the blue line in the inset in Fig. \ref{fig:5}(b).  As $n^o_1$ increases, guided DSWs disappear entirely when $n^o_1=n^e_2(90 \dgr)=2$, resulting in $\phi_{2c}=-\Delta_{cL}=90\dgr$, as the index corresponding to the radiation channel becomes the highest refractive index in the system. In this process, the SBICs move towards the cutoff of the DSW and also disappear. It is however possible to recover the line of SBICs by tuning the other constitutive parameters of the system \cite{Mukherjee2019}. This is shown in Figs. \ref{fig:5}(c) and (d), where in addition to $n^o_1=n^e_2(90 \dgr)=2$, we set $n_1^e=1.2$. Since $\Re(N)<n^o_1$ for all the range of existence of DSWs and the radiation channel is always accessible, the structure only supports leaky DSWs so that the line of SBICs, shown in the inset in Fig. \ref{fig:5}(d), are the only bound surface modes supported by the system. The leaky DSW mode becomes more radiative at higher values of $\Delta$ as shown in Fig.~\ref{fig:5}(b) and (d).

\subsection{Materials}

To show the feasibility of experimentally observing the SBICs described above, we first note that DSW modes have been experimentally observed, for example, at the interface of nematic liquid crystals \cite{Li2020}. Thus, we consider the interface between the liquid crystal E7 in the nematic phase and a calcite crystal, arranged so that $\Delta=-50\dgr$. We consider two different wavelengths $\lambda_0$: 488 $nm$ and 632 $nm$. At $\lambda_0=488\ nm$, the refractive indices of the materials are $n_{E7}^o=1.5345, n_{E7}^e=1.7754, n_{calcite}^o=1.6674, n_{calcite}^e=1.4904$ \cite{Li2005, Ghosh1999}, and the structure supports both guided and leaky DSWs in the range of optic axis orientation $50.65\dgr<\phi_2^{DSW}<51.48\dgr$. Within the leaky DSW range, there exists a SBIC at $\phi_2^{BIC}=50.72\dgr$. Due to material dispersion, the refractive indices of the materials are slightly different at $\lambda_0=632\ nm$ with $n_{E7}^o=1.5189, n_{E7}^e=1.7304, n_{calcite}^o=1.6557, n_{calcite}^e=1.4849$. As a result, such structure supports leaky DSWs in the range $55.39\dgr<\phi_2^{DSW}<56.05\dgr$ with an embedded SBIC at $\phi_2^{BIC}=55.95\dgr$. Therefore,  material dispersion leads to the two situations discussed in Fig.~\ref{fig:5}, with the SBIC occurring at different values of $\phi_2$ for each wavelength. Therefore, on the one hand, one concludes that material dispersion affects the type and location in the parameter space of the surface states supported by the structure, but SBICs do exist. On the other hand, intrinsic material absorption is not relevant in this context, because the considered materials are highly transparent and anyway BICs are exclusively related to the suppression of radiative losses.  As comparison, the absorption coefficient for E7 at $\lambda_0=632\ nm$ is $0.03\ cm^{-1}$ \cite{Wu1987}, resulting in a propagation length $L_{E7} \sim 30\ cm$, which is orders of magnitude larger than the propagation length of the leaky DSW due to radiation losses, which typically is in the range $L_{leaky} \sim 10-100\ \mu m$. 

\section{Conclusions}

In summary, we have uncovered the existence of SBICs  realized as Dyakonov states in uniaxial anisotropic media, in structures where Dyakonov modes were considered to be impossible.  To the best of our knowledge, they are the first example of SBICs supported by the interface between two homogeneous media.  The existence loci of the new states is set  by the amount of anisotropy-asymmetry and the constitutive parameters of the materials. Here we addressed only azimuthal anisotropy-symmetry breaking and uniaxial media, but the Dyakonov mechanism occurs in more general geometries and types of anisotropy, where we anticipate that richer families of states may exist. Our results open the possibility to create SBICs and Dyakonov-like states in a whole new class of materials and metamaterials, including different types of anisotropic materials, such as hyperbolic materials. 

Finally, it has to be properly appreciated that the SBICs we found here are anisotropy-induced \cite{Gomis-Bresco2017}. Namely, they are full vector, hybrid surface states situated above the light line and exist without the presence of a trapping potential. Therefore, they do not arise from the interference of resonances, as, e.g., Friedrich–Wintgen BICs do. Rather, the SBICs occur when the ordinarily-polarized wave that constitutes the radiation channel is not needed to fulfill the boundary conditions. This mechanism makes the anisotropic SBICs fundamentally different from all surface BICs known to date \cite{Molina2012, Weimann2013, Corrielli2013, Gallo2014, Hsu2013a, Hu2017}, in particular those where a defect located at the interface acts as a trapping potential.

\vspace{.5cm}
This work was partially supported by the H2020 Marie Skłodowska-Curie Action GA665884; Grants CEX2019-000910-S and PGC2018-097035-B-I00 funded by MCIN/AEI/10.13039/501100011033/FEDER, Fundació Cellex, Fundació Mir-Puig, and Generalitat de Catalunya (CERCA). 

\bibliography{references}

\end{document}